To be submitted to Microscopy and Microanalysis

# Composition-Driven Phase Evolution in Sm-Doped $BiFeO_3$ via Latent-Field Reconstruction of Atomically Resolved STEM Data

Newsha Javanmardi[1]*, Christopher T. Nelson[2], Anna N. Morozovska[3], Eugene A. Eliseev[4], Ichiro Takeuchi[5], and Sergei V. Kalinin[1]*

[1] Department of Materials Science and Engineering, University of Tennessee, Knoxville, Tennessee 37996, USA
[2] Center for Nanophase Materials Sciences, Oak Ridge National Laboratory, Oak Ridge, Tennessee 37831, USA
[3] Institute of Physics, National Academy of Sciences of Ukraine, 46 Nauky Avenue, Kyiv 03028, Ukraine
[4] Frantsevich Institute for Problems of Materials Science, National Academy of Sciences of Ukraine, 3 Omeliana Pritsaka Street, Kyiv 03142, Ukraine
[5] Department of Materials Science and Engineering, University of Maryland, College Park, Maryland 20742, USA

* Corresponding authors: Newsha Javanmardi and Sergei V. Kalinin; sergei2@utk.edu

Functionalities of ferroelectric materials are governed by the spatial organization and coupling of polarization, strain, lattice rotation, and structural order accessible via atomically resolved scanning transmission electron microscopy (STEM) images. Quantitative interpretation of atomic-resolution STEM data has conventionally relied on locating atomic columns and converting their fitted coordinates into local structural descriptors. Here, we develop a field-based approach in which atomic-resolution images are represented by spatially varying latent Bragg fields, whose amplitudes and phases provide continuous maps of crystalline order, lattice displacement, strain, rotation, and mode-specific residual structure. The observed atomically resolved images are decoded from the latent fields. We apply this framework to image series of Sm-substituted $BiFeO_3$ spanning 0-20% Sm and crossing the composition-driven boundary between the R3c ferroelectric phase and the orthorhombic, nonpolar Pnma phase. Conventional atom-resolved parameterization is used as an independent validation, showing that reconstructed Bragg amplitude tracks local atomic-column intensity and that field-derived shear reproduces unit-cell angular distortions obtained from atom fitting. The combined analysis reveals a systematic evolution from extended ferroelectric domains at low Sm concentration, through the appearance and growth of localized regions with period-doubled Pnma order at intermediate compositions, to a connected Pnma-dominated state at high Sm content. The period-doubled order is accompanied by enhanced shear and lattice rotation and by progressive reorganization of the ferroelectric domain structure. These results establish latent-field reconstruction as a physically interpretable complement to atom finding and provide a unified framework for resolving composition-driven phase evolution in ferroic materials.

## I. Introduction

Ferroelectrics are a materials family in which symmetry breaking is associated with the emergence of spontaneous electric polarization that can be reoriented by an applied electric field.[1] In conventional proper ferroelectrics, polarization is the primary order parameter, but its spatial organization and response cannot be understood independently of the other structural and electrostatic degrees of freedom.[2] Polarization couples to strain through electrostriction and to electric fields through electrostatic energy.[3] It also couples to chemical composition, charged defects, surfaces, interfaces, and, in many perovskites, oxygen-octahedral rotations and tilts.[4] These interactions establish characteristic length scales extending from atomic displacements within a unit cell to domain walls and mesoscale domain patterns.[5] All of these structures contribute to macroscopic electromechanical response.[6] Consequently, ferroelectric behavior is determined not only by the symmetry or lattice parameters of an average structure, but also by the spatial distribution of local order parameters and the compatibility conditions that couple them.

This multiscale character becomes especially pronounced in thin films, heterostructures, compositionally complex materials, and systems near structural phase transition boundaries.[2] Epitaxial strain can stabilize phase states absent in bulk crystals and produce strain-driven morphotropic-like boundaries.[7] Electrostatic and chemical boundary conditions can stabilize conductive domain walls [8] and chemically reconstructed interfaces.[9] Related ferroic heterostructures host flux-closure quadrants,[10] polar vortices,[11] and skyrmions.[12] In these systems, nominally small changes in local lattice distortion may correspond to qualitatively different physical states. Establishing structure-property relationships therefore requires methods capable of identifying not only the average phase but also spatial variations in polarization, strain, rotation, chemical site occupancy, and interfacial structure.

Aberration-corrected STEM provides direct access to projected atomic-column positions and intensities with picometer-scale precision.[13] In perovskite ferroelectrics, displacement of a transition-metal or B-site cation relative to the surrounding A-site sublattice can be used as a local structural proxy for polarization.[13] Measurements of local lattice vectors provide tetragonality, shear, rotation, and strain, whereas column-shape and oxygen-sensitive imaging can make octahedral distortions accessible.[14] Interface-sensitive STEM measurements can relate structural distortions to charge compensation[15] and polarization-controlled screening.[9] These capabilities enabled unit-cell-scale mapping of ferroelectricity and tetragonality,[13] determination of dipole configurations at charged and uncharged walls,[16] mapping of coupled polarization and octahedral tilts,[14] and observation of emergent polar textures.[10]

However, an atomic-resolution STEM image is not itself a direct map of polarization or any other thermodynamic field. It is a two-dimensional projection generated by electron scattering through a specimen of finite thickness and is influenced by probe shape, crystal tilt, scan distortions, drift, dose, and noise.[17],[18] The structural polarization inferred from projected cation displacements is also distinct from a rigorous three-dimensional polarization calculated using the complete ionic configuration and the corresponding Born effective charges. Moreover, displacement fields of physical interest are often on the order of only a few picometers, comparable to the apparent displacements that can be introduced by scan instabilities or imperfect image registration.[19],[20] Quantitative interpretation therefore depends as strongly on the image-analysis pathway as on the nominal spatial resolution of the microscope.

The first step in most atomic-resolution analyses is correction of instrumental distortions and establishment of a reliable coordinate system. Revolving or orthogonal scan acquisition can suppress drift-induced distortions.[20],[21] Non-rigid registration improves frame averaging and removes time-dependent image deformation.[22],[23] These approaches enable picometer-precision coordinate measurements when combined with appropriate uncertainty controls.[19] Once corrected, the image can be analyzed through pathways that differ in the quantities assumed to be observable, the structural priors imposed, and whether the output is a deterministic estimate or a probability distribution over possible structures.

The most direct pathway is real-space atom finding. In this case, atomic columns are identified as local intensity maxima and their positions are refined using Gaussian, centroid, or template-based fitting.[24],[25] The resulting coordinates can be indexed by sublattice or unit cell and converted into local lattice vectors, bond lengths, bond angles, cation displacements, strain tensors, and polarization proxies.[13] This representation is physically intuitive and highly effective when atomic columns are well separated, image contrast is strong, and local lattice topology is known. Its limitations become important in low-signal regions, at interfaces and defects, in overlapping projections, and in strongly distorted or multiphase materials, where errors in peak assignment propagate nonlinearly into subsequently calculated fields.

A second pathway operates in reciprocal space. Geometric phase analysis isolates selected Bragg components and uses their spatially varying phases to determine displacement and strain relative to a reference lattice.[26] Related peak-pair methods provide a real-space route to local strain measurement.[27] These approaches are naturally sensitive to periodic order and can yield smoothly varying displacement fields without explicitly locating every atom. However, the result depends on the selected reciprocal-lattice vectors, Fourier masks, reference region, and assumptions concerning phase continuity. Fourier filtering can also suppress nonperiodic information associated with interfaces, defects, chemistry, and local motif changes. Real-space atom fitting and Fourier-phase analysis should therefore be viewed as complementary projections of the image rather than interchangeable measurements.

A third class of methods seeks to infer local structure directly from image neighborhoods or learned representations. Local-crystallography approaches identify phases, symmetries, and defects from atomic neighborhoods.[28] Deep-learning methods can identify chemistry and track local transformations,[29] while statistical representations can extract physically relevant descriptors from atomically resolved images.[30] Bayesian analysis enables uncertainty-aware comparison of physical models for ferroelectric domain walls.[31] Polarization distributions can be learned with or without an explicit atom-finding intermediate.[32] Four-dimensional STEM retains a diffraction pattern at every probe position and extends the available reciprocal-space observables.[33] Electron ptychography provides a related reconstructive route to phase and scattering-potential information.[34]

Despite this progress, most current analysis workflows remain sequential. Images are first corrected, atomic columns are then fitted, atomic coordinates are converted into displacement or polarization vectors, and these vectors are subsequently interpreted using a physical model. Each stage produces a point estimate that is treated as exact by the next stage, even though the underlying quantities may be incompletely observed or mutually coupled. Fourier-based workflows similarly require the reference lattice and relevant Bragg modes to be specified before the physical fields are reconstructed. These choices are frequently reasonable, but they obscure the relationship between assumptions, observability, and uncertainty. They are especially restrictive for systems containing multiple phases, spatially varying strain, diffuse interfaces, weak or missing atomic columns, and coupled polar and nonpolar structural modes.

Here, we propose a field-first reconstruction in which atomic-resolution images are represented by physically interpretable latent Bragg fields and the observed images are decoded from these latent fields. We apply the method to image series of Sm-doped $BiFeO_3$ spanning the R3c-to-Pnma composition series. Conventional atom coordinates and unit-cell descriptors independently validate the reconstructed amplitude and deformation fields and provide motif-level polar-displacement and Pnma-related quantities not determined by the Bragg fields alone. We combine the two representations to follow the composition-dependent emergence, growth, and connectivity of period-doubled Pnma order across the morphotropic boundary.

## II. Latent Field Analysis

Quantitative analysis of atomic-resolution STEM images is commonly formulated as a sequence of object-detection operations: atomic columns are identified, their coordinates and intensities are refined, atoms are assigned to crystallographic sublattices, and local structural descriptors are calculated from the resulting coordinate set. This representation has enabled unit-cell-scale measurements of ferroelectric displacements, lattice distortions, strain, and polarization-related order and remains the conventional reference pathway for atomic-resolution ferroelectric microscopy.[13] Here, we adopt a complementary field-based representation in which the image is treated as an observation generated by a small number of spatially modulated crystalline modes.

### II.A. Field-based representations of atomic structures

The fundamental variables of the analysis are the continuous complex fields rather than discrete atomic coordinates. For an image $I(\mathbf{r})$, where $\mathbf{r} = (x, y)$ denotes position in the image plane, we write the observation model as

$$I(\mathbf{r}) = B(\mathbf{r}) + \sum_{j=1}^{N_q} \mathrm{Re}\left[\eta_j(\mathbf{r})\exp\left(i\mathbf{q}_j \cdot \mathbf{r}\right)\right] + \epsilon(\mathbf{r}). \quad (1)$$

Here, $B(\mathbf{r})$ represents a slowly varying background, $\mathbf{q}_j$ are reference reciprocal-lattice vectors, $\eta_j(\mathbf{r})$ are slowly varying complex latent fields, and $\epsilon(\mathbf{r})$ describes image noise and structural components not represented by the selected modes. Each latent field is written as

$$\eta_j(\mathbf{r}) = A_j(\mathbf{r})\exp\left[i\phi_j(\mathbf{r})\right], \quad (2)$$

where $A_j(\mathbf{r})$ is the local mode amplitude and $\phi_j(\mathbf{r})$ is its phase. This form is closely related to geometric-phase and local lock-in descriptions of crystalline images,[26] but is used here as a reconstructive latent representation rather than only as a strain-measurement transform.

The decomposition separates two physically distinct forms of image variation. Changes in $A_j$ reflect changes in the strength and coherence of the corresponding periodic image component and can arise from local atomic-column intensity, structural disorder, thickness, occupancy, defects, or loss of crystalline coherence. Variations in $\phi_j$, by contrast, describe local translations of periodic lattice relative to the reference state. When several non-collinear reciprocal vectors are available, the phase fields jointly define a two-dimensional displacement field and, through its spatial derivatives, local strain, and lattice rotation.

The representation is latent because the complex fields are not measured directly by the detector. They are inferred from the image under the assumption that atomic-resolution contrast can be represented locally by modulated crystalline modes. Unlike a generic learned latent vector, these fields retain an explicit relation to reciprocal-space structure, lattice displacement, and continuum deformation. They therefore provide a physically constrained intermediate representation between the raw image and atom-derived structural descriptors. This complements local-crystallography methods[28] and machine-learning approaches that extract phases, motifs, and transformations from atomic-resolution images.[29]

### II.B. Identification and demodulation of crystalline modes

The images were first restricted to the measured contrast-bearing support, and a missing-aware low-pass background was removed before Fourier analysis. Full preprocessing, cropping, masking, and parameter settings are provided in the Supplementary Methods and in the accompanying analysis notebook. The resulting high-pass image is

$$I_{\mathrm{hp}}(\mathbf{r}) = I(\mathbf{r}) - B(\mathbf{r}). \tag{3}$$

The resulting high-pass image contains the periodic lattice contrast used for Bragg-mode identification and complex demodulation. Candidate reciprocal vectors are identified from local maxima in the magnitude of the Fourier transform,

$$\tilde{I}(\mathbf{q}) = \mathcal{F}\left[I_{\mathrm{hp}}(\mathbf{r})\right]. \tag{4}$$

The central low-frequency region is excluded, and candidate peaks are ranked by intensity, separation from the origin, angular diversity, and consistency with approximately orthogonal or symmetry-related lattice directions. For the two-dimensional displacement reconstruction used here, at least two non-collinear reciprocal vectors are required.

For each selected reciprocal vector $\mathbf{q}_j$, the image is shifted to baseband and low-pass filtered:

$$\eta_j(\mathbf{r}) = G_{\sigma_j} * \left[I_{\mathrm{hp}}(\mathbf{r})\exp\left(-i\mathbf{q}_j \cdot \mathbf{r}\right)\right], \tag{5}$$

where $G_{\sigma_j}$ is a Gaussian envelope and the asterisk denotes convolution. This operation is equivalent to a local lock-in measurement of the crystalline component associated with $\mathbf{q}_j$. The filter width controls the compromise between spatial resolution and reciprocal-space selectivity. A narrow real-space kernel retains sharp interfaces but mixes nearby reciprocal components, whereas a broad kernel isolates the Bragg mode more cleanly but smooths domain walls and localized defects.

For image contrasts that exhibit an approximate sign or sublattice ambiguity, the raw phase can be doubled before unwrapping. The doubled phase is unwrapped, divided by two, and corrected for a fitted affine background. Removal of the affine component eliminates the arbitrary global phase, constant translation, and small mismatch between the selected reciprocal vectors and the mean experimental lattice:

$$\phi_j^{\mathrm{res}}(\mathbf{r}) = \phi_j(\mathbf{r}) - \left(a_j x + b_j y + c_j\right). \tag{6}$$

The residual phase then contains local deviations from the best-fit reference lattice.

### II.C. Displacement, strain, and rotation fields

For a smoothly displaced lattice, the phase of the $j$th crystalline mode is related to the displacement field $\mathbf{u}(\mathbf{r})$ by

$$\phi_j^{\mathrm{res}}(\mathbf{r}) \approx -\mathbf{q}_j \cdot \mathbf{u}(\mathbf{r}). \tag{7}$$

Writing the reciprocal vectors as rows of a matrix,

$$\mathbf{Q} = \begin{bmatrix} q_{1x} & q_{1y} \\ q_{2x} & q_{2y} \\ \vdots & \vdots \end{bmatrix}, \tag{8}$$

and the residual phases as

$$\boldsymbol{\Phi} = \begin{bmatrix} \phi_1^{\mathrm{res}} \\ \phi_2^{\mathrm{res}} \\ \vdots \end{bmatrix}, \tag{9}$$

the displacement is obtained through an amplitude-weighted least-squares inversion,

$$\mathbf{u}(\mathbf{r}) = -[\mathbf{Q}^{\mathrm{T}}\mathbf{W}(\mathbf{r})\mathbf{Q}]^{-1}\mathbf{Q}^{\mathrm{T}}\mathbf{W}(\mathbf{r})\boldsymbol{\phi}(\mathbf{r}). \tag{10}$$

The diagonal matrix $\mathbf{W}(\mathbf{r})$ weights each phase equation according to the local Bragg amplitude. Modes with weak local amplitude therefore contribute less strongly to the displacement estimate, reducing phase instabilities in defects, interfaces, and low-signal regions.

The displacement field is smoothed using an amplitude-weighted local estimator before differentiation. Its gradient is decomposed into symmetric and antisymmetric parts:

$$\nabla\mathbf{u} = \begin{bmatrix} \partial_x u_x & \partial_y u_x \\ \partial_x u_y & \partial_y u_y \end{bmatrix}, \tag{11}$$

$$\boldsymbol{\varepsilon} = \frac{1}{2}[\nabla\mathbf{u} + (\nabla\mathbf{u})^{\mathrm{T}}], \qquad \boldsymbol{\omega} = \frac{1}{2}[\nabla\mathbf{u} - (\nabla\mathbf{u})^{\mathrm{T}}]. \tag{12}$$

The corresponding in-plane components are

$$\varepsilon_{xx} = \frac{\partial u_x}{\partial x}, \qquad \varepsilon_{yy} = \frac{\partial u_y}{\partial y}, \tag{13}$$

$$\varepsilon_{xy} = \frac{1}{2}\left(\frac{\partial u_x}{\partial y} + \frac{\partial u_y}{\partial x}\right), \qquad \omega = \frac{1}{2}\left(\frac{\partial u_y}{\partial x} - \frac{\partial u_x}{\partial y}\right). \tag{14}$$

Because differentiation amplifies high-frequency noise, the strain and rotation fields are more sensitive than displacement to scan-line artifacts, phase-unwrapping errors, missing support, and filter selection. Quantitative interpretation is therefore restricted to regions with sufficient Bragg amplitude and adequate distance from crop boundaries and missing-data masks.

### II.D. Residual phases and image reconstruction

The displacement field accounts for phase variation that can be represented as a common translation of the selected lattice modes. The remaining mode-specific phase,

$$\chi_j(\mathbf{r}) = \phi_j^{\mathrm{res}}(\mathbf{r}) + \mathbf{q}_j \cdot \mathbf{u}(\mathbf{r}), \tag{15}$$

contains information not captured by a single displacement field. This residual can arise from local changes in the crystallographic basis or motif, phase-slip structures, coexistence of structural variants, errors in the selected reference lattice, or image components outside the assumed observation model.

Because $\chi_j$ is angular, ordinary linear correlations with the wrapped phase are not well defined. It is represented through its circular components,

$$s_j(\mathbf{r}) = \sin\chi_j(\mathbf{r}), \qquad c_j(\mathbf{r}) = \cos\chi_j(\mathbf{r}), \tag{16}$$

which preserve the periodic character of the variable and permit comparison with other local descriptors. The adequacy of the latent representation is evaluated by reconstructing the image:

$$I_{\mathrm{rec}}(\mathbf{r}) = B(\mathbf{r}) + \sum_j \mathrm{Re}\left[\eta_j(\mathbf{r})\exp\left(i\mathbf{q}_j \cdot \mathbf{r}\right)\right]. \tag{17}$$

The reconstruction residual,

$$R(\mathbf{r}) = I(\mathbf{r}) - I_{\mathrm{rec}}(\mathbf{r}), \tag{18}$$

provides a measure of model incompleteness. A small residual indicates that the selected modes reproduce the dominant periodic contrast but does not establish that every derived field is physically correct. Conversely, spatially structured residuals can identify defects, nonperiodic motif variations, interfaces, or additional reciprocal modes that require a richer representation.

### II.E. Validation against atom-resolved parameterization

The latent-field reconstruction does not require atom finding for its construction. Conventional atom-resolved parameterization is used both as an independent validation representation and as the source of motif-level quantities that are not uniquely available from the Bragg fields. Bayesian analysis of atomically resolved ferroelectric domain-wall data established the use of unit-cell descriptors as inputs to physical inference.[31] The same Sm-doped $BiFeO_3$ composition series was subsequently analyzed through causal relations among local structural, chemical, and polarization descriptors.[35] Polarization distributions from this system were also learned both with and without an explicit atom-finding stage.[32] Here, atomic-column coordinates and intensities are grouped into local unit cells, from which cation displacement, lattice parameters, unit-cell angle, area, and intensity descriptors are calculated.

The atom-derived unit-cell centers are registered to the cropped image without resizing either representation. Because microscopy files can store coordinates in Cartesian $(x, y)$ order or array $(\mathrm{row}, \mathrm{column})$ order, both conventions are evaluated using image-bound occupancy and agreement between local image intensity and atom-derived intensity. The selected coordinate transformation is applied consistently to cropping, interpolation, visualization, and correlation analysis. Latent fields are sampled at registered unit-cell centers. Cells outside the measured image support, within the missing-data buffer, or close to the crop boundary are excluded. Pearson and Spearman correlations are then calculated between image-derived latent fields and atom-derived descriptors. This comparison tests whether a field reconstruction inferred independently of atomic coordinates recovers physically meaningful structural variations. The most direct expected correspondences are between local Bragg amplitude and atomic-column intensity, and between field-derived shear and atom-derived unit-cell angular distortion.

The validation also defines the limits of the representation. Strong agreement for amplitude and shear does not imply that every motif-level degree of freedom can be inferred from a small set of Bragg modes. Rather, it establishes which physical quantities are observable in the selected field representation and which require additional structural information.

### II.F. Relation between lattice displacement and polarization

The displacement field reconstructed from Bragg phases describes translation and deformation of the selected periodic lattice. It is not generally equivalent to ferroelectric polarization. Structural polarization in a perovskite is associated with relative displacements of different sublattices within the unit cell, whereas the Bragg-phase displacement describes motion of the periodic image motif as a whole. Methods that infer polarization directly from image neighborhoods without an explicit atom-finding stage nevertheless require training labels or an equivalent structural reference.[32] Schematically, $u_{\mathrm{Bragg}}$ is not equal to P.

The measured atom-derived polar displacement is more appropriately written as

$$\mathbf{P}_{\mathrm{app}}(\mathbf{r}) = \mathbf{P}_{\mathrm{struct}}(\mathbf{r}) + \mathbf{b}_{\mathrm{tilt}}(\mathbf{r}) + \mathbf{b}_{\mathrm{scan}}(\mathbf{r}) + \boldsymbol{\epsilon}(\mathbf{r}). \quad (19)$$

Here, $\mathbf{b}_{\mathrm{tilt}}$ includes apparent displacement generated by specimen mistilt, thickness-dependent channeling, and dynamical scattering, while $\mathbf{b}_{\mathrm{scan}}$ represents residual instrumental distortion. These common-mode contributions can produce a nonzero apparent polarization even in a weakly polar or nonpolar structure. [17],[18],[32]

For this reason, atom-derived polarization is analyzed separately from the Bragg displacement reconstruction. Established unit-cell analyses infer polarization-related order from relative sublattice displacements[13] and probabilistic analyses treat these local descriptors as distinct observables.[31] Polarization-vector populations are modeled in the $(P_x, P_y)$ plane, and an inversion-center or common-mode offset is estimated when symmetry-related domain variants provide sufficient information. The corrected displacement is

$$\mathbf{P}_{\mathrm{corr}}(\mathbf{r}) = \mathbf{P}_{\mathrm{app}}(\mathbf{r}) - \mathbf{b}_{\mathrm{common}}(\mathbf{r}). \quad (20)$$

When the data contains only one domain variant and no independent nonpolar reference, common-mode bias and true polarization are not uniquely identifiable. In such cases, the analysis reports apparent polar displacement and marks the correction as weakly constrained rather than assigning an absolute polarization magnitude.

This separation establishes a hierarchy of observables. Bragg amplitudes describe the strength of periodic order; Bragg phases describe average-lattice displacement and deformation; residual mode structure describes deviations from the common displacement model; and motif-level relative displacements are required to determine polarization and antipolar order.

### II.G. Opportunities and limitations of the latent-field representation

The latent-field approach does not require every atomic column to be detected and assigned before a displacement field can be constructed. It retains a continuous representation across the image, permits direct reconstruction of the observation, and provides a common framework in which amplitude, displacement, strain, rotation, residual phase, and phase identity can be analyzed jointly. It is also naturally extensible to probabilistic inference, where the latent fields and model parameters are represented by posterior distributions rather than single optimized values.

The representation nevertheless depends on the selected reciprocal vectors, demodulation length scales, reference-lattice gauge, and validity of the slowly varying envelope approximation. It can be affected by scan distortion, specimen bending, local changes in thickness, missing data, and overlap of neighboring Fourier components. A small number of Bragg modes cannot uniquely determine all motif-level degrees of freedom. Polarization, oxygen-octahedral rotations, chemical order, and antipolar displacements can require additional modes or an explicit crystallographic-basis model.

The present formulation should therefore be viewed as complementary to atom-resolved analysis rather than as a universal replacement. Atom finding provides direct local crystallographic measurements when the motif is known and well resolved. Latent-field reconstruction provides a continuous, reconstructive description that is advantageous for extended deformation, phase coexistence, missing regions, and structural patterns whose natural description is field-like rather than atom-discrete. The same conceptual framework can be expanded to 4D-STEM, where reciprocal-space information is retained at every probe position[33] and additional latent fields can be constrained by nanodiffraction or ptychographic observables.[34]

## III. Phase evolution in Sm-doped $BiFeO_3$

Samarium substitution provides a compositionally controlled path from the rhombohedral polar R3c phase of $BiFeO_3$ to an orthorhombic antipolar or non-ferroelectric Pnma phase. The morphotropic boundary occurs near 14% Sm and is accompanied by enhanced dielectric and electromechanical response.[36] Electron diffraction showed short-range antiparallel cation displacements below the boundary, a complex nanoscale phase mixture at the critical composition, and orthorhombic twin structures at Sm concentrations above the boundary.[37] The high-Sm Pnma state is associated with antipolar A-site displacements and an octahedral-tilt superstructure that introduces half-order reflections and a doubled projected periodicity.[37] Near the boundary, additional quarter-order and incommensurate superstructure components reveal a modulated Pnma-related bridging state rather than a spatially uniform direct R3c-to-Pnma transformation.[37] In projected angle-distortion and shear maps, this period-doubled order appears as alternating stripe-like bands. The stripe-like appearance is descriptive only; throughout this paper, the scientific designation is the period-doubled Pnma phase or, for the near-boundary intermediate state, period-doubled Pnma-related order.

Synchrotron diffraction established coexistence of competing structures near the boundary[38] and composition-dependent measurements mapped polarization rotation[39] and structural evolution with temperature and rare-earth substitution.[40] Phenomenological modeling connected the polar-to-antipolar transition to shallow competition between the R3c and Pnma states,[41] while microstructure-electromechanical measurements linked the boundary morphology to enhanced response.[42]

The present dataset contains 14 paired HAADF-STEM and unit-cell-parameterization datasets at nominal Sm concentrations of 0, 7, 10, 13, and 20%.[43] Earlier atom-resolved analysis of this system revealed spatially modulated order-parameter fields near the ferroelectric-antiferroelectric boundary and attributed their stabilization to flexoelectric coupling.[44] The same composition series was later used for causal analysis of competing atomistic mechanisms[35] and for learning polarization distributions with and without atom finding.[32] Here, the conventional descriptors validate the field reconstruction and provide the motif-level quantities used to identify Pnma-related colonies and ferroelectric walls. The colony score is obtained from directional band-pass structure in the atom-derived angle-distortion field using one threshold set at the pooled 0% Sm 98.5th percentile. Ferroelectric walls are obtained from the mistilt-corrected apparent polar-displacement direction, and their charged or uncharged labels use a geometric bound-charge proxy.

### III.A. Continuous fields in the rhombohedral R3c ferroelectric phase

The 0% Sm-free dataset provides the reference R3c ferroelectric state (Figure 1). The Fourier transform contains a sharp and nearly orthogonal set of lattice reflections, and the selected Bragg modes reconstruct the dominant atomic contrast with $R^2 = 0.951$ for the high-pass lattice contrast and 0.955 for the full cropped image (NRMSE = 0.211). The reconstruction residual is concentrated near the support boundary and the localized interior defect, whereas the latent amplitude is comparatively uniform throughout the crystalline region. This establishes that a small set of complex fields captures the periodic lattice contrast without requiring atomic-column localization.

The reconstructed displacement components vary smoothly over the field of view and contain broad regions that coincide qualitatively with atom-derived polar-displacement domains. This correspondence should not be interpreted as equality between Bragg displacement and polarization: the former describes translation of the selected periodic lattice, whereas the latter is obtained from relative sublattice displacement. The atom-derived angle-distortion map contains only weak fluctuations and isolated defect-associated features, and neither the angle field nor the latent shear displays extended Pnma-related period doubling. The derivative fields do contain a diagonal line feature and high-frequency scan texture, demonstrating why local strain and rotation must be interpreted together with the amplitude mask, residual image, and composition controls.

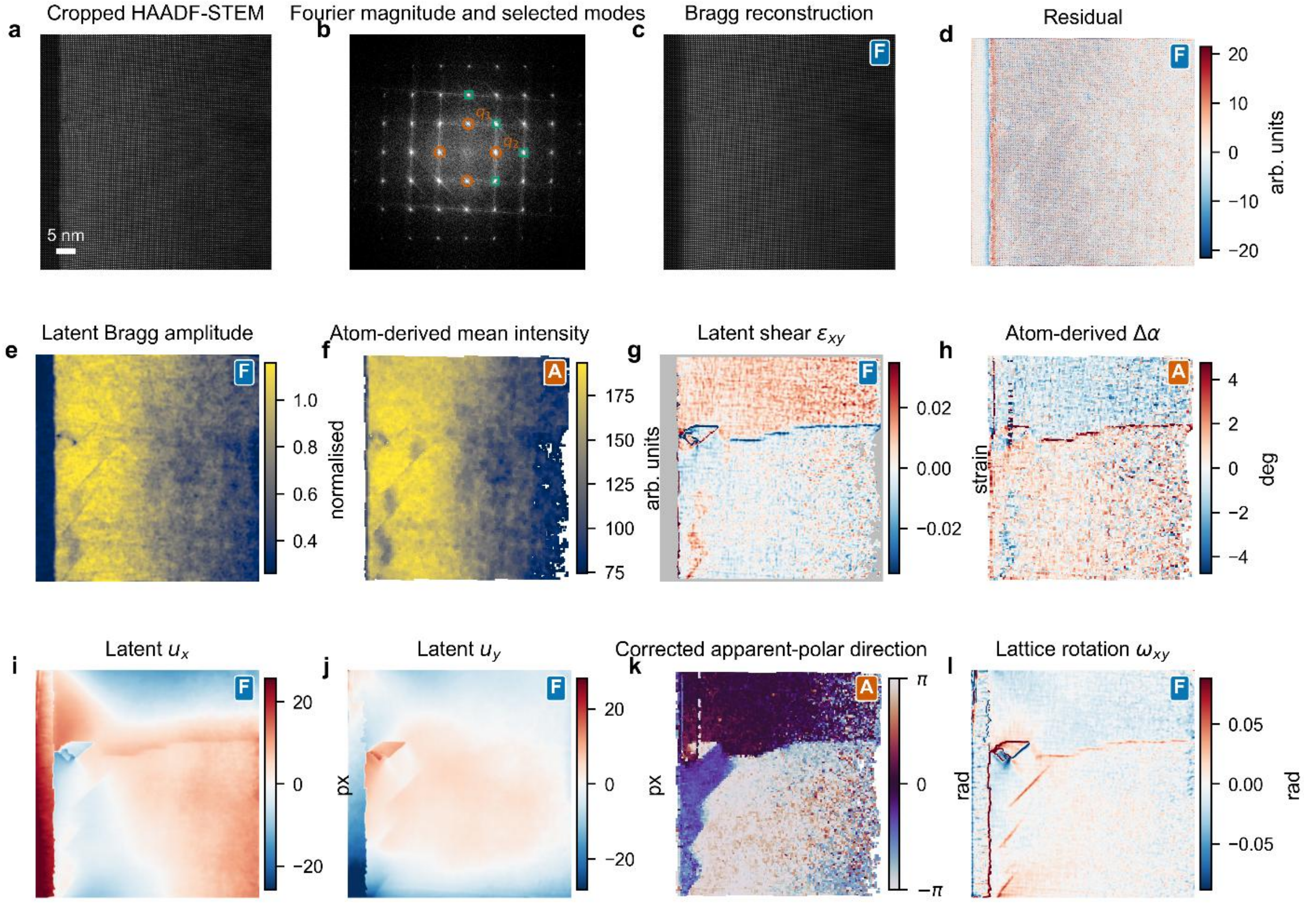

**Figure 1.** Latent-field reconstruction and conventional atom-based quantities for Sm_0_1 (0% Sm). F denotes latent-field analysis and A denotes conventional atom-column analysis used for validation; the input and Fourier-selection diagnostic are unbadged. (a) Cropped HAADF-STEM image; only the analyzed crop is shown. (b) Fourier magnitude; circles mark the two primary non-collinear reciprocal vectors and squares mark additional reconstruction modes. (c) Bragg reconstruction and (d) residual. (e, f) Latent Bragg amplitude and the corresponding atom-derived mean column intensity. (g, h) Field-derived shear and atom-derived unit-cell angle distortion. All shear maps in Figures 1–3 are displayed within quantitative support on a common −0.035 to +0.035 strain scale; gray indicates unsupported pixels. (i, j) In-plane latent displacement components. (k) Mistilt-corrected apparent polar-displacement direction and (l) lattice rotation. The panel order and real-space orientation are identical in Figures 1 and 2.

### III.B. Continuous fields near the morphotropic phase boundary

A representative 13% Sm dataset is shown in Figure 2. This composition lies immediately below the nominal 14% morphotropic boundary and exhibits coexistence between slowly varying R3c-like lattice deformation and a localized region of period-doubled Pnma-related order. The selected Bragg fields reproduce the dominant image contrast with $R^2 = 0.953$ for the high-pass lattice contrast and 0.974 for the full cropped image (NRMSE = 0.161). The latent amplitude remains continuous across most of the image, while the displacement fields contain broad gradients together with a periodic component in the lower part of the field of view.

The period-doubled Pnma-related region is most clearly identified by alternating positive and negative unit-cell angle distortion. The same spatial region appears independently in field-derived shear and rotation, demonstrating that the superstructure is encoded in the image phase rather than introduced solely by atom finding. Atom-derived polar-displacement components also reorganize across this region, but their absolute magnitude is not used because a common mistilt or channeling contribution can shift the apparent displacement vector. The central observation is therefore the co-localization of atom-derived angular modulation with independently reconstructed shear and rotation.

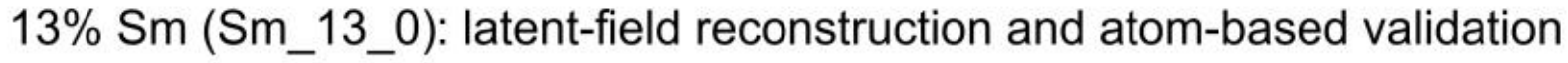

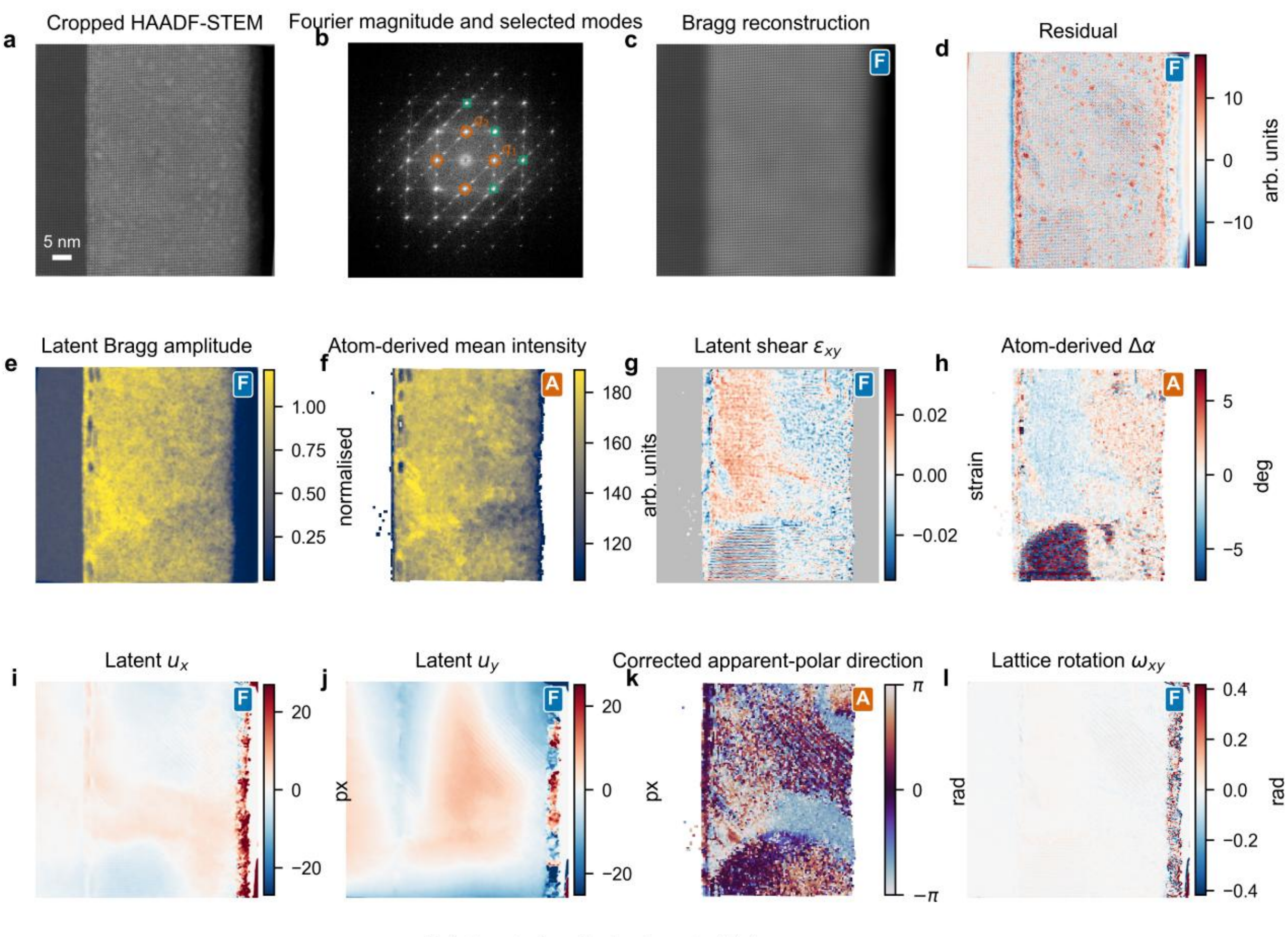

**Figure 2.** Latent-field reconstruction and conventional atom-based quantities for Sm_13_0 (13% Sm). F denotes latent-field analysis and A denotes conventional atom-column analysis used for validation; the input and Fourier-selection diagnostic are unbadged. (a) Cropped HAADF-STEM image; only the analyzed crop is shown. (b) Fourier magnitude; circles mark the two primary non-collinear reciprocal vectors and squares mark additional reconstruction modes. (c) Bragg reconstruction and (d) residual. (e, f) Latent Bragg amplitude and the corresponding atom-derived mean column intensity. (g, h) Field-derived shear and atom-derived unit-cell angle distortion. All shear maps in Figures 1–3 are displayed within quantitative support on a common −0.035 to +0.035 strain scale; gray indicates unsupported pixels. (i, j) In-plane latent displacement components. (k) Mistilt-corrected apparent polar-displacement direction and (l) lattice rotation. The panel order and real-space orientation are identical in Figures 1 and 2.

The validation was performed over all 14 datasets rather than only for the representative 13% image. Fisher-weighted repository correlations show that latent Bragg amplitude tracks atom-derived mean column intensity (Pearson $r = 0.844$) and column-intensity dispersion ($r = 0.752$). The geometric validation is similarly strong: field-derived $\varepsilon_{yy}$ correlates with atom-derived a-axis strain at $r = 0.852$, $\varepsilon_{xx}$ correlates with b-axis strain at $r = 0.795$, and field-derived shear correlates with unit-cell angle distortion at $r = -0.834$ (95% interval −0.913 to −0.694). The shear sign follows the stored coordinate convention; its magnitude and consistency show that the phase-derived deformation tensor recovers the same local angular distortion measured by conventional unit-cell geometry. Bragg displacement and atom-derived polar displacement show weaker and less consistent component-wise correlations, as expected for observables defined at different structural levels.

### III.C. Composition-dependent evolution of the period-doubled Pnma phase

Figure 3 compares representative images at all five compositions in a fixed-orientation four-row layout. The first row shows the direction of the continuous latent lattice-displacement field, the second row shows latent shear within the quantitative-support mask on one common scale, the third row shows detected period-doubled Pnma-related colonies, and the fourth row shows the relationship of those colonies to ferroelectric walls classified by the geometric bound-charge proxy. The undoped R3c image has only sparse detector responses; the 7 and 10% images contain isolated, replicate-dependent colonies; the 13% image contains a coherent colony; and the 20% image is dominated by one connected colony. Charged and uncharged wall segments both occur near colony boundaries, so the comparison supports local coexistence rather than a universal charged-wall nucleation rule.

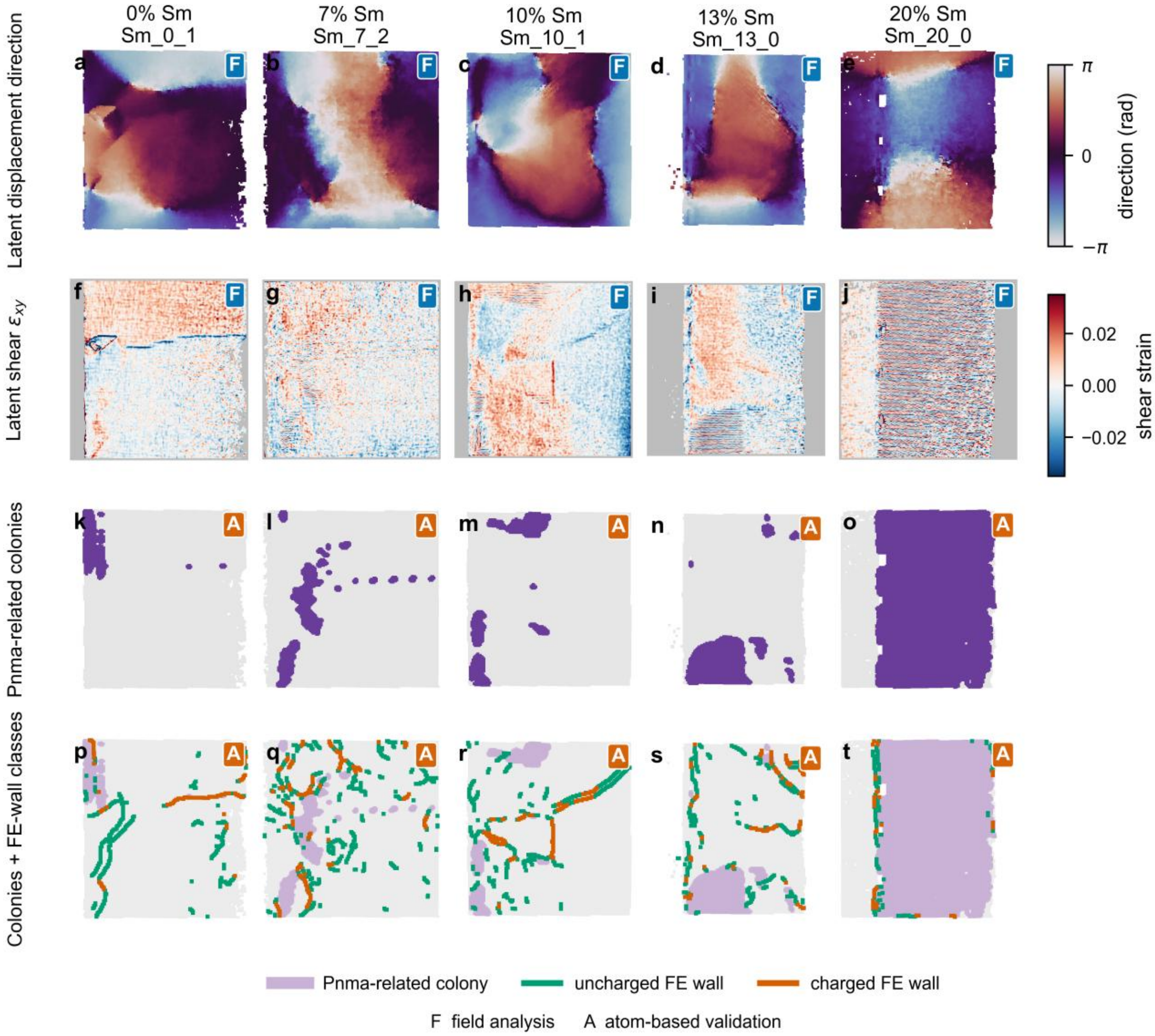

**Figure 3.** Composition-dependent comparison of latent-field results and conventional atom-based validation for representative 0, 7, 10, 13, and 20% Sm images. F denotes latent-field analysis and A denotes conventional atom-column analysis used for validation. (a–e) Direction of the continuous Bragg-phase displacement field; this relative lattice-translation field is not polarization. (f–j) Latent shear within the quantitative-support mask. All five shear maps use the same −0.035 to +0.035 strain scale used in Figures 1 and 2; gray indicates unsupported pixels. (k–o) Detected period-doubled Pnma-related colonies. (p–t) Pnma-related colony mask with the ferroelectric wall skeleton overlaid; green and orange identify walls classified as uncharged and charged, respectively. All maps use stored crop coordinates, so orientation is identical within each column and directly comparable with Figures 1 and 2. The wall charge label is a sign-based structural proxy, not a calibrated bound-charge density.

The composition-level metrics are summarized in Figure 4 using the stored full-resolution arrays. The mean detected Pnma-related area fractions are 0.015 ± 0.019 at 0% Sm, 0.067 ± 0.029

at 7%, 0.032 ± 0.043 at 10%, 0.180 ± 0.053 at 13%, and 0.665 ± 0.123 at 20%. The small nonzero low-composition values are treated as a detector baseline rather than a physical bulk Pnma fraction. The robust change is the increase between 10 and 13% Sm, followed by formation of an extended Pnma-dominated region at 20% Sm. The ferroelectric wall-density proxy increases from 0.0305 ± 0.0090 in the undoped material to 0.0629 ± 0.0164 at 20% Sm, indicating progressive reorganization of the slowly varying polar-displacement field.

Within detected Pnma-related regions, the modulation period is approximately two pseudocubic unit cells at 7, 10, 13, and 20% Sm. The 0% estimate of 2.33 ± 0.58 cells is produced by sparse detector responses and is not interpreted as a bulk R3c modulation. Composition therefore changes the area and connectivity of the Pnma-related phase much more strongly than its characteristic wavelength. The twofold diffuse-FFT anisotropy A2 is not monotonic and has large replicate scatter at 20% Sm; the complete A2, A4, angular-streak, and spectral-entropy comparison is retained in Figure S4. Global reciprocal-space streaking cannot by itself distinguish the Pnma superstructure from scan-axis artifacts, making the local real-space correspondence between angle distortion and shear the more reliable signature.

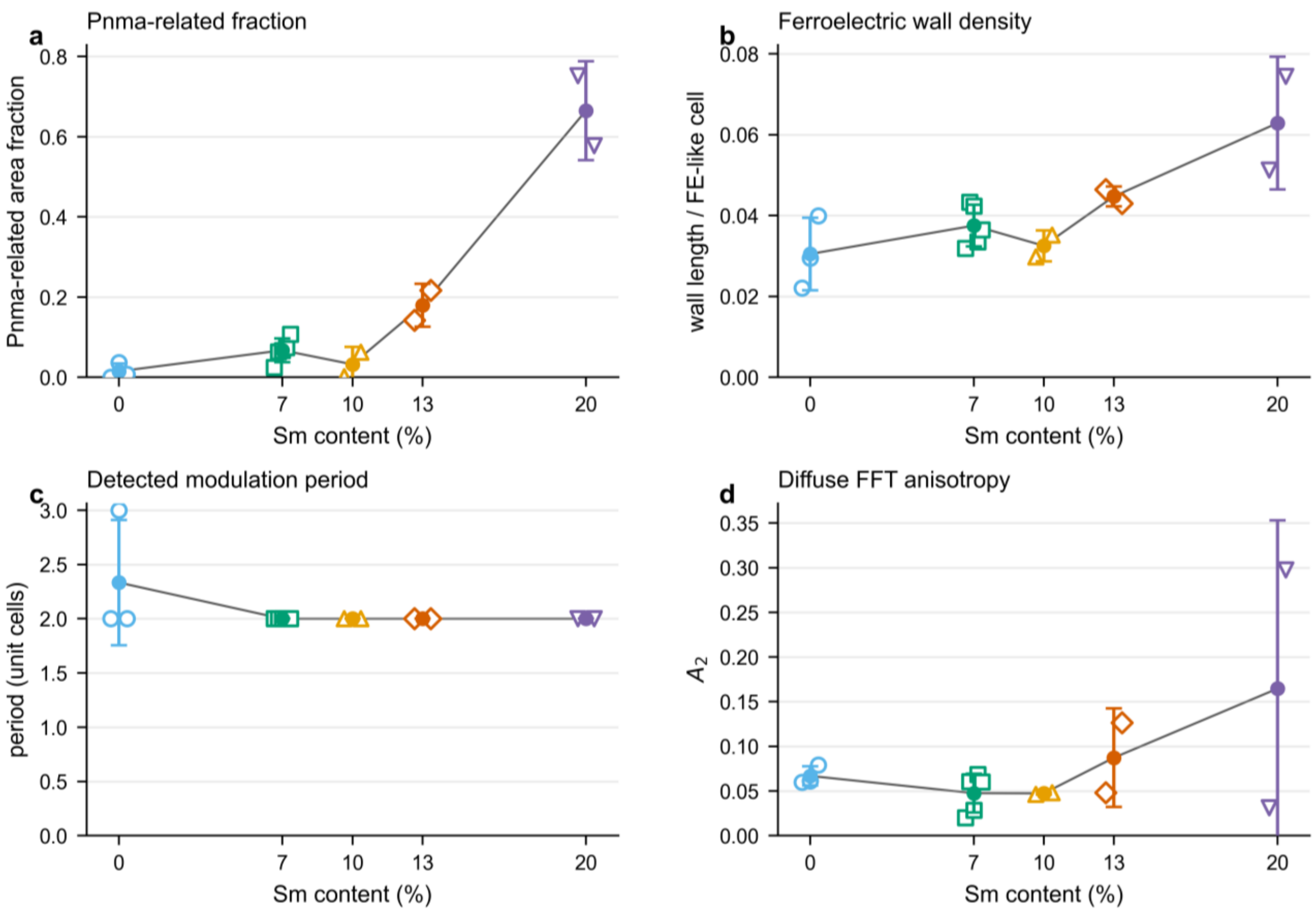

**Figure 4.** Composition-dependent metrics computed from the stored full-resolution arrays for all 14 images. Open symbols are individual images; filled symbols and error bars are the composition mean and standard deviation between images. (a) Detected period-doubled Pnma-related area fraction. (b) Ferroelectric wall length per ferroelectric-like unit cell. (c) Modulation period within detected Pnma-related regions. (d) Twofold anisotropy A2 of the Bragg-excluded diffuse Fourier intensity.

### III.D. Nucleation, growth, and coalescence of period-doubled Pnma order

The combined field and metric analysis define a simple morphology sequence. At low Sm concentration, the material is dominated by extended R3c ferroelectric domains and only sparse detector responses to Pnma-related modulation. At 7–10% Sm, localized period-doubled Pnma regions appear, but their fraction is small and varies substantially between images. Near 13% Sm, these regions develop into coherent Pnma-related colonies. By 20% Sm, the detected Pnma-related area fraction is $0.665 \pm 0.123$ and the largest connected component occupies $0.664 \pm 0.125$ of the usable field, showing that nearly all detected Pnma-related area has coalesced into one component. The transition is therefore best described as nucleation, growth, and coalescence of period-doubled Pnma order rather than as a continuous reduction of modulation wavelength.

This evolution is consistent with earlier diffraction and atom-resolved observations of Pnma-related modulated phases near the ferroelectric-antiferroelectric boundary.[37] In the earlier atomic-scale analysis, the near-boundary phase displayed modulated structural and polarization order parameters, and flexoelectric coupling was proposed to reduce the effective wall energy and stabilize the modulated state. [44] The present analysis does not independently determine whether flexoelectricity, elastic compatibility, electrostatic boundary conditions, or chemical heterogeneity supplies the dominant microscopic stabilization. It does show that the Pnma-related superstructure is simultaneously encoded in atom-derived unit-cell geometry and continuous Bragg-phase fields, and that its principal composition-dependent change is growth and connectivity together with an increasing ferroelectric wall-density proxy.

## IV. Summary

Methodologically, we developed a field-first reconstruction of atomically resolved STEM images in which a small set of complex Bragg envelopes generates continuous maps of crystalline amplitude, displacement, strain, rotation, and residual phase. The reconstruction is performed without atomic coordinates, while conventional unit-cell parameterization provides independent validation and the motif-level Pnma and polar descriptors not uniquely available from the Bragg fields. Agreement between latent amplitude and atom-column intensity, and between phase-derived shear and unit-cell angular distortion, identifies the physical content of the latent representation and clarifies which motif-level quantities, particularly polarization, still require sublattice-resolved analysis.

For the Sm-doped $BiFeO_3$ series, the combined latent-field and atom-resolved analysis reveals systematic evolution from an undoped R3c ferroelectric state without coherent period doubling, through sparse and replicate-dependent Pnma-related nuclei at 7–10% Sm, to coherent period-doubled Pnma colonies near 13% Sm and a connected Pnma-dominated state at 20% Sm. The mean detected Pnma-related fraction increases sharply between 10 and 13% Sm and reaches $0.665 \pm 0.123$ at 20% Sm. Once established, the modulation period remains approximately two unit cells, whereas the ferroelectric wall-density proxy increases and the period-doubled angle field becomes strongly co-localized with shear and rotation. The composition dependence is therefore governed primarily by the area and connectivity of Pnma-related order rather than by a continuously changing modulation wavelength.

The practical advantage of the latent representation is most apparent relative to peak-by-peak atom finding and in machine-learning-enabled microscopy. Peak fitting remains indispensable when sublattice identity and absolute motif geometry are required, but it is brittle when columns are missing, overlapping, weak, or locally rearranged. Continuous latent fields use all image pixels, preserve spatial continuity, provide physically interpretable channels for learning and classification, and can be evaluated even where an atomic list is incomplete. They therefore offer a natural intermediate representation for automated microscopy: more directly connected to image formation than a table of fitted coordinates, but substantially more interpretable and transferable than an unconstrained neural-network embedding. Recent diffraction-based work by Latypov and co-workers represents crystalline heterogeneity through spatially resolved latent descriptors. [45] A complementary continuous-field formulation treats diffraction intensity as a differentiable field for reconstructing patterns and localizing boundaries and heterogeneity. [46]

## V. Acknowledgements

This material (NJ, IT, SVK) is based upon work supported by the National Science Foundation under Award No. NSF 2523284.

The work of A.N.M. and E.A.E. is primarily supported by the DOE Software Project on "Computational Mesoscale Science and Open Software for Quantum Materials," under Award Number DE-SC0020145 as part of the Computational Materials Sciences Program of US Department of Energy, Office of Science, Basic Energy Sciences, and also partially supported by the National Academy of Sciences of Ukraine.